\documentclass[aps,prl,twocolumn,superscriptaddress]{revtex4}

\usepackage{graphicx}
\usepackage{dcolumn}
\usepackage{bm}
\usepackage{color}

\begin{document}
\title{Low-Temperature Collapse of Electron Localisation in Two Dimensions}
\author{Matthias Baenninger}
\email{matthias.baenninger@cantab.net}
\affiliation{Cavendish Laboratory, University
of Cambridge, J.J. Thomson Avenue, Cambridge CB3 0HE, United
Kingdom.}
\affiliation{Department of
Physics, Indian Institute of Science, Bangalore 560 012, India.}
\author{Arindam Ghosh}
\email{arindam@physics.iisc.ernet.in}
\affiliation{Department of
Physics, Indian Institute of Science, Bangalore 560 012, India.}
\author{Michael Pepper}
\affiliation{Cavendish Laboratory, University of Cambridge, J.J.
Thomson Avenue, Cambridge CB3 0HE, United Kingdom.}
\author{Harvey E. Beere}
\affiliation{Cavendish Laboratory, University of Cambridge, J.J.
Thomson Avenue, Cambridge CB3 0HE, United Kingdom.}
\author{Ian Farrer}
\affiliation{Cavendish Laboratory, University of Cambridge, J.J.
Thomson Avenue, Cambridge CB3 0HE, United Kingdom.}
\author{David A. Ritchie}
\affiliation{Cavendish Laboratory, University of Cambridge, J.J.
Thomson Avenue, Cambridge CB3 0HE, United Kingdom.}
\date{\today}

\begin{abstract}
We report direct experimental evidence that the insulating phase of a disordered, yet strongly interacting two-dimensional electron system (2DES) becomes unstable at low temperatures.  As the temperature decreases, a transition from insulating to metal-like transport behaviour is observed, which persists even when the resistivity
of the system greatly exceeds the quantum of resistivity $h/e^2$. The results have been achieved by measuring transport on a mesoscopic length-scale while systematically varying the strength of disorder.
\end{abstract}
\maketitle
Ever since the first two-dimensional electron systems (2DES) were realised, they have been used extensively to investigate various theories and concepts of charge localisation~\cite{Mott1975}. The scaling theory of localisation prohibits extended electronic states in two dimensions (2D) at absolute zero in presence of disorder~\cite{Abrahams1979}. While there has been extensive theoretical work on the possibility of a delocalisation caused by electron-electron interactions, e.g.~\cite{Abrahams2001,Fleishman1980,Shepelyansky1994,Basko2006,Benenti1999,Katomeris2003,Henseler2007}, conclusive experimental evidence of such an effect has not been observed. On the contrary, the insulating phase in 2D at low temperatures has proven robust, in particular in the case of strong localisation, where the resistivity $\rho\gg h/e^2$.

For strong disorder, insulating variable-range hopping transport has been observed, with interaction effects only leading to a modification of the single-particle density of states~\cite{Efros1975}. In low but finite disorder an interaction driven localisation mechanism has been suggested in form of pinned charge-density waves~\cite{Tanatar1989, Normand1992}. However, no deviation from the insulating nature of transport has been reported in potential realisations of such phases, nor is it expected theoretically.  The intermediate regime where disorder and interaction effects are equally important, is very challenging to study theoretically and experimentally and is still not well understood. Our work represents an attempt to close this gap.

A crucial property of disorder is the characteristic length-scale of its potential fluctuations. If the disorder is mainly long-range, at low electron densities the system becomes increasingly inhomogeneous. Transport then behaves according to classical percolation law, masking possible interactions between electrons~\cite{Sarma2005}. Hence, an experimental approach for investigating interaction effects in presence of disorder should minimise the effects of long-range disorder and focus on short-range fluctuations.

In modulation doped GaAs/AlGaAs heterojunctions, the disorder mainly comes from the remote charged ions in the doping layer, and the strength of disorder depends strongly on the width $\delta$ of the undoped spacer layer between 2DES and doping layer. The possibility of changing the strength of disorder by varying $\delta$ provides a powerful tool in the investigation of disorder effects. In theoretical treatment, an entirely random distribution of the dopants is generally assumed, giving an uniform spectral density of the potential fluctuations with only fluctuations of length-scale $<\delta$ exponentially damped in the 2DES~\cite{Efros1988}. However, recent imaging of the disorder landscape of 2DES suggest that the dominant length scale in modulation doped GaAs/AlGaAs heterostructures is, in fact, greater than 0.5 $\mu$m $\gg \delta$~\cite{Finkelstein2000, Chakraborty2004}. This strongly indicates that in these systems, long-range disorder dominates on a macroscopic length-scale.

In consideration of this problem, our experimental approach differs in two crucial ways from previous studies of charge localisation in 2D:  (1) We have used mesoscopic 2DES extending only over 0.5 $\mu$m to a few microns, thereby strongly reducing the impact of long-range disorder and allowing a focus on the short-range fluctuations of order $\delta$ (schematic in Fig.~1b). (2) Instead of outright minimisation, we have systematically varied the strength of background
potential fluctuations by varying the spacer width. On the other hand, the electron density $n_{\rm s}$ and, hence, interaction parameter $r_{\rm s}=1/a_{\rm B}^{*}\sqrt{\pi n_{\rm s}}$ ($a_{\rm B}^{*}$ the effective Bohr radius), could be tuned with a metallic top gate.
Devices were fabricated from Si monolayer doped GaAs/AlGaAs heterojunctions, where the spacer between 2DES and doping layer was varied from 20~nm to 60~nm with a total depth of the 2DES 270~nm$-$300~nm. A typical doping concentration of $n_{\delta}=2.5\times 10^{12}$~cm$^{-2}$ was used and as-grown mobilities were $0.6-1.8\times10^{6}$~cm$^{2}/$Vs. The dimensions of the 2DES were determined by a wet-etched mesa of width $W\sim 8~\mu$m and the length $L\sim 0.5-3~\mu$m  of a NiCrAu top gate, which was defined by optical or e-beam lithography. A typical set of devices is shown in Fig.~1a.

We have recently confirmed that localisation in macroscopic devices, indeed, shows characteristics of a percolation transition caused by long-range charge inhomogeneities~\cite{Baenninger2005}. This percolation behaviour is absent in mesoscopic devices from the same wafer. Additionally, freeze-out of transport occurs at much lower densities, making a new regime of disorder and interaction strength accessible. In this regime, a completely new behaviour of the $T$-dependence of resistance in the localised regime was observed, which is the focus of this letter.

In Figs.~1c-e we show typical temperature traces for one macroscopic and two mesoscopic devices with approximately the same resistivity at lowest $T$. All the devices had the same spacer $\delta=40$ nm. The macroscopic device shows the expected activated behaviour down to the lowest temperatures.  In contrast, the mesoscopic devices show activated transport only at high $T$, followed by a temperature range where the resistivity increases only slowly or even decreases as $T$ is lowered.  In the following, we use the terms ``metallic'' for a positive temperature coefficient of resistivity $d\rho/dT>0$ and ``saturated'' for the weak but non-metallic $T$-dependence, as opposed to ``insulating'' for an activated behaviour. Great care was taken to rule out electron heating as the origin of the saturated $T$-dependence, even though heating could, of course, not explain the metallic behaviour. The main results presented here were obtained in a dilution refrigerator with base mixing chamber temperature $T_{\rm{Base}}\cong60$~mK. The electron temperature was separately estimated as $\approx70$~mK using a Coulomb blockade thermometer. Measurements were done in a four-probe set-up with appropriate radiation filtering and a constant excitation voltage $V_{\rm{ext}}\cong5~\mu$V$<k_{\rm{B}}T_{\rm{Base}}/e$ and frequency $\Omega_{\rm{ext}}=84$~Hz, where current and voltage drop across the sample were measured simultaneously. Both saturated and metallic behaviour were also observed in a $^{3}$He cryostat with $T_{\rm{Base}}\cong300$~mK, where a four-probe constant current set-up was used ($I_{\rm{ext}}$=0.1~nA, $\Omega_{\rm{ext}}=7.3$~Hz). In both systems, the $T$-dependent damping of Shubnikov-de Haas oscillations showed no saturation of the electron temperature at lowest $T$. The low choice of excitation voltage or current ensures that experiments were within the linear response regime, which was confirmed by a separate measurement. Note that the metallic behaviour also rules out a finite size effect as origin of the unexpected low temperature behaviour.
 
Figs.~2a-d show $\rho$ plotted against $1/T$ at four values of $n_{\rm s}$ marked by arrows
in the inset to 2a for a mesoscopic device with particularly strong metallic behaviour. We focus on the following distinctive features: (1) At all $n_{\rm s}$, metallic behaviour appears at low $T$, while the
system shows insulator-like activated transport above a crossover temperature $T_{\theta}$, indicating a temperature-driven insulator-to-metal transition. (2) The metallic behaviour can be detected over
nearly three orders of magnitude of $\rho$ and up to $\rho\approx 700~h/e^2$. 
These observations clearly distinguish our results from previously reported
examples of zero-temperature quantum phase transitions in 2D,
including superconductor-to-insulator~\cite{Haviland1989}, quantum Hall liquid-to-insulator~\cite{Shahar1995} or apparent metal-insulator transitions~\cite{Abrahams2001, Simmons1998}, all of which occur at a scale set by $\rho \sim h/e^2$. $T_{\theta}$ varies from 0.35~K to 1.6~K, which is up to one order of magnitude lower than the corresponding Fermi temperature (3.7-7~K), implying our systems to be degenerate. The resistivity as a function of $n_{\rm s}$  is shown in the inset of Fig.~2a for three values of $T = 60$~mK, 500~mK, and 4.2~K. The trace at 500~mK shows
highest $\rho$ at all $n_{\rm{s}}$ ($0.9-1.7\times10^{10}$~cm$^{-2}$), confirming a non-monotonic $T-$dependence over the
entire accessible localised regime.

In order to highlight the generality of the phenomenon we show in Fig.~3 the ``metallicity'' (expressed by the slope $\frac{d\rho(T)/dT}{\rho(T)}$ at $T=300$ mK) for four devices from three different wafers with spacer $\delta$=20-60 nm. All devices had a mesa width of 8~$\mu$m, but the length of the gate varied from $L\sim 0.5~\mu$m  to $3~\mu$m. The colour coding discerns the regions of $n_{\rm{s}}$ with a metallic $T$-dependence from those with saturated behaviour. Device III, on which this letter is focused, shows a continuous metallic phase over the whole density range while other devices show both metallic and saturated regions. Extent of metallicity varies from device to device and also between different cooldowns of the same device, but more than 50\% of the mesoscopic devices with $20$~nm$\leq\delta\leq60$~nm showed metallic behaviour at some windows of electron density. Importantly, neither saturated nor metallic $T$-dependence were observed in a mesoscopic device with $\delta=80$~nm. On the other hand, for $\delta\leq10$~nm Coulomb blockade effects often dominated the $T$-dependence~\cite{Ghosh2004a}. This highlights the importance of finding the ideal window of disorder  strength. In all cases, the metallic behaviour could be suppressed by applying a
perpendicular magnetic field $B_{\perp} \gtrsim 1 - 1.5$~T, but a transition to a saturated regime at low $T$ persisted. One example is shown in the inset of Fig.~2d.

As the low-temperature behaviour observed in our devices is not expected for conventional one-particle Anderson localisation~\cite{Mott1975} or tunnelling between electron droplets~\cite{Tripathi2006}, alternative mechanisms have to be considered. There have been various suggestions how interaction effects could affect the localisation in disordered 2DES. Enhanced screening or formation of many-body states has been suggested to lead to extended or less localised wave functions in the groundstate or above a critical temperature, and modify the single-electron conduction mechanism~\cite{Fleishman1980,Shepelyansky1994,Basko2006,Benenti1999,Henseler2007,Slutskin2000}. On the other hand, the interaction effects could also be suppressed at high temperatures, which would explain our observed behaviour qualitatively, with $T_{\theta}$ as the temperature below which the delocalisation sets in. A renewed localisation at even lower temperatures cannot be ruled out, although no sign of such a second turnaround has been observed experimentally.

A possible description may be based on the behaviour of defects in an interaction-induced, disorder stabilised pinned electron quantum solid (QS): The existence of delocalised zero-point defects (defectons) in a crystal with strong zero-point fluctuations was first proposed for solid helium~\cite{Andreev1969} and has been adapted for 2DES~\cite{Shapiro2001, Spivak2003, Katomeris2003}. Moderate disorder has been predicted to facilitate the formation of a Wigner crystal for $r_{\rm s}$ close to the values in our 2DES ($r_{\rm{s}}\approx 4-6$)~\cite{Chui1995}, a range where defecton formation is likely~\cite{Baenninger2005}. Suppression of the metallic state by a perpendicular magnetic field is then explained by localisation of these charged (quasi)-particles at local potential traps. Since, here, these traps arise from the distribution of the electrons themselves, transport then occurs by a weakly $T$-dependent, near-resonant tunnelling of localised defects over the average distance of a lattice constant.
Similarly, disorder can induce an energy gradient between defect states. If the energy difference exceeds the defecton bandwidth, delocalisation is suppressed and transport again occurs by tunnelling of localised defects. Hence, the duality of metallic and saturated transport is explained by the influence of the local disorder potential, which determines the extension of the metallic phase.
A particularly important aspect is that analysis of the magnetoresistance in the regime of saturated $T$-dependence revealed an universal average hopping distance $\lambda_{\rm{hop}}\approx r_{\rm{ee}}\cong1/\sqrt{n_{\rm s}}$, as expected for defect tunnelling in an electron solid~\cite{Ghosh2004, Baenninger2005}. This observation also excludes a spin entropy effect as a likely origin of the metallic behaviour~\cite{Spivak2003}.

In analogy to atomic transport in solid $^3$He, and noting that the atomic diffusion coefficient is analogous to the inverse resistivity in the
electronic QS, we can write the total $\rho(T)$ in the form~\cite{Pushkarov2003}
\begin{equation}
\label{eq1} \rho(T)^{-1} = (\rho_0 + \alpha T^\gamma)^{-1} +
\beta \exp\left[-E_0/k_{\rm B}T\right].
\end{equation}
\noindent The second term represents the
hopping transport of localised defects, which dominates at $T>T_{\theta}$, with $E_0$ the
energy barrier. The first term signifies quantum diffusion of
mobile defectons with $\rho_0$ the
residual resistivity. In Figs.~2a-d we have successfully fitted Eq.~\ref{eq1} to our data. For the temperature exponent we found $\gamma \approx 2$ (Fig.~4b). This suggests inter-quasiparticle scattering
in a gas of degenerate defectons as the dominant relaxation mechanism~\cite{Pushkarov2003}. The deviations from $\gamma \
\approx 2$ at lowest densities may be explained by onset of stronger disorder scattering, making it harder to resolve the $T^2$-component of the resistivity. The activation energy was found to be $E_{0}/k_{\rm B}\approx1-6$K, with highest $E_{0}$ at lowest $n_{\rm s}$ (not shown).

The fitting parameters are relevant not only with respect to defecton transport but also in a more general context. The observed $T^{2}$-dependence of $\rho$ at low $T$ is expected in the standard Fermi liquid theory for strongly interacting metals. The finite resistance $\rho_{0}$ extrapolated to $T=0$ (Fig.~4a) fits into this picture, however, for the case of a metal, $\rho_{0}\lesssim h/e^{2}$ is expected, contrary to our observation $\rho_{0}\gg h/e^{2}$. Phenomenologically, $T_{\theta}$ indicates the crossover temperature between insulating and metallic state, and, thus, defines a phase diagram in $T-n_{\rm s}$-space (Fig.~4c). $T_{\theta}$ shows a non-monotonic behaviour with a maximum for intermediate $n_{\rm s}$. Neither the unusually high values of $\rho_{0}$, nor the behaviour of $T_{\theta}$ are understood at present.
Furthermore, the question remains if the metallic $T$-dependence is the signature of a  metallic ground state in 2D or if it will disappear at lower temperatures due to the onset of quantum corrections~\cite{Simmons2000}.

We thank EPSRC for funding. M.B. acknowledges the support by
the EU funded COLLECT network, the Sunburst Trust, and the Cambridge Overseas Trust.

\begin{figure*}[h]
\centering
\includegraphics[width=0.6\textwidth]{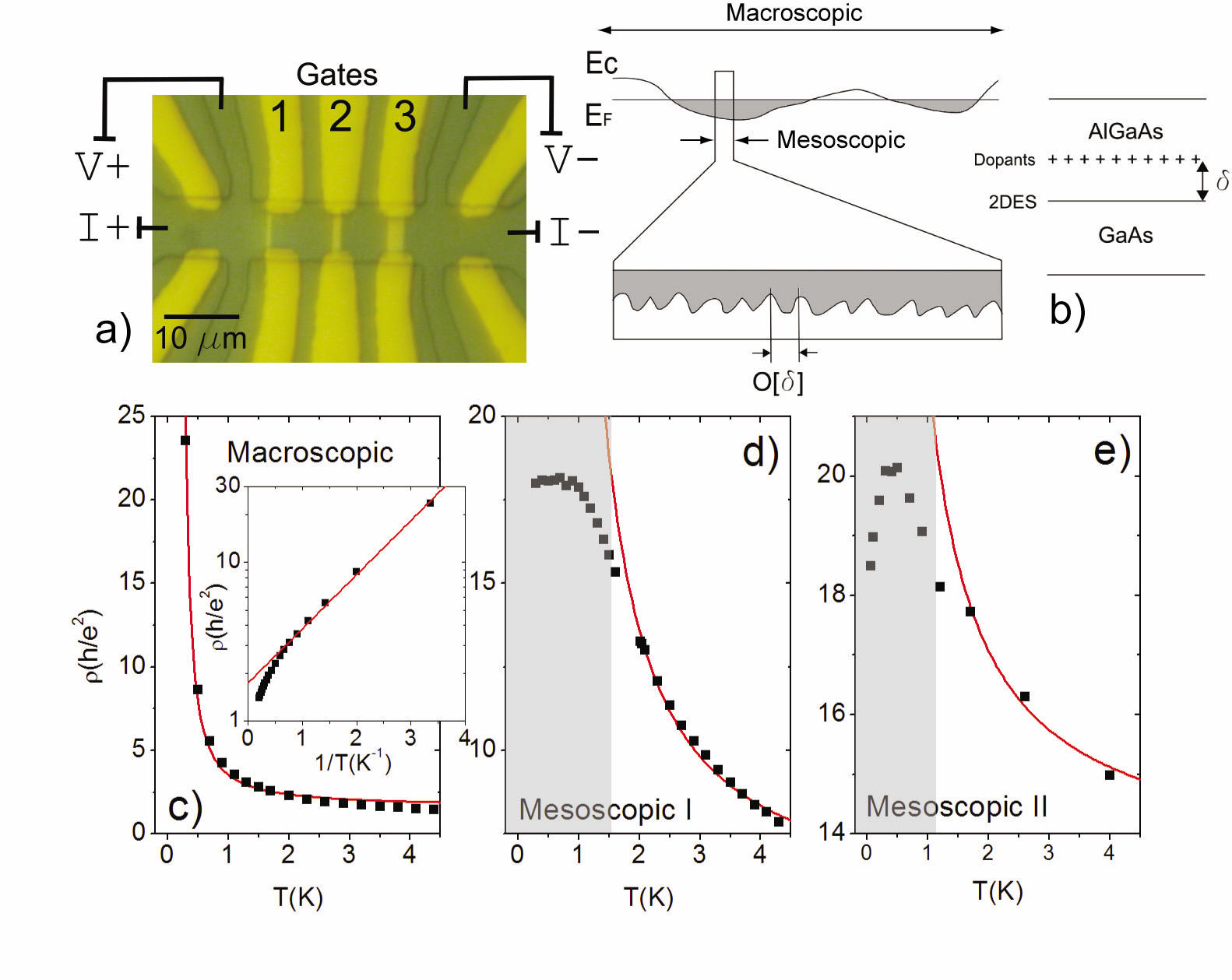}
\caption{(Colour online) a) Optical microscope picture of a typical device with mesa width $W=8~\mu$m. The gates labelled 1-3 have lengths $L$=0.5,$ $ 1, and $2~\mu$m. The other gates were not used. b) Schematic representation of long- and short-range disorder in mesoscopic devices. For devices smaller than the length scale of long-range disorder, short-range fluctuations of $O[\delta]$ are dominant.
c)-e) Comparison of temperature dependence of resistivity for two mesoscopic devices ($W=8~\mu$m, $L=2,~0.5~\mu\rm m$ for device I, II) with a macroscopic one ($100~\mu\rm m \times 900~\mu$m) from wafers with the same spacer (40~nm). While the macroscopic device shows an exponential increase in resistivity to lowest $T$, a saturation or downturn occurs in the mesoscopic devices. Solid lines are fits of the form $\rho=\rho_{0}\exp\left[E_0/k_{\rm B}T\right]$. The inset to c) shows the same data on a $\log(\rho)$ against 1/$T$ scale, highlighting the exponential behaviour.}
\end{figure*}

\begin{figure*}[h]
\centering
\includegraphics[width=0.6\textwidth]{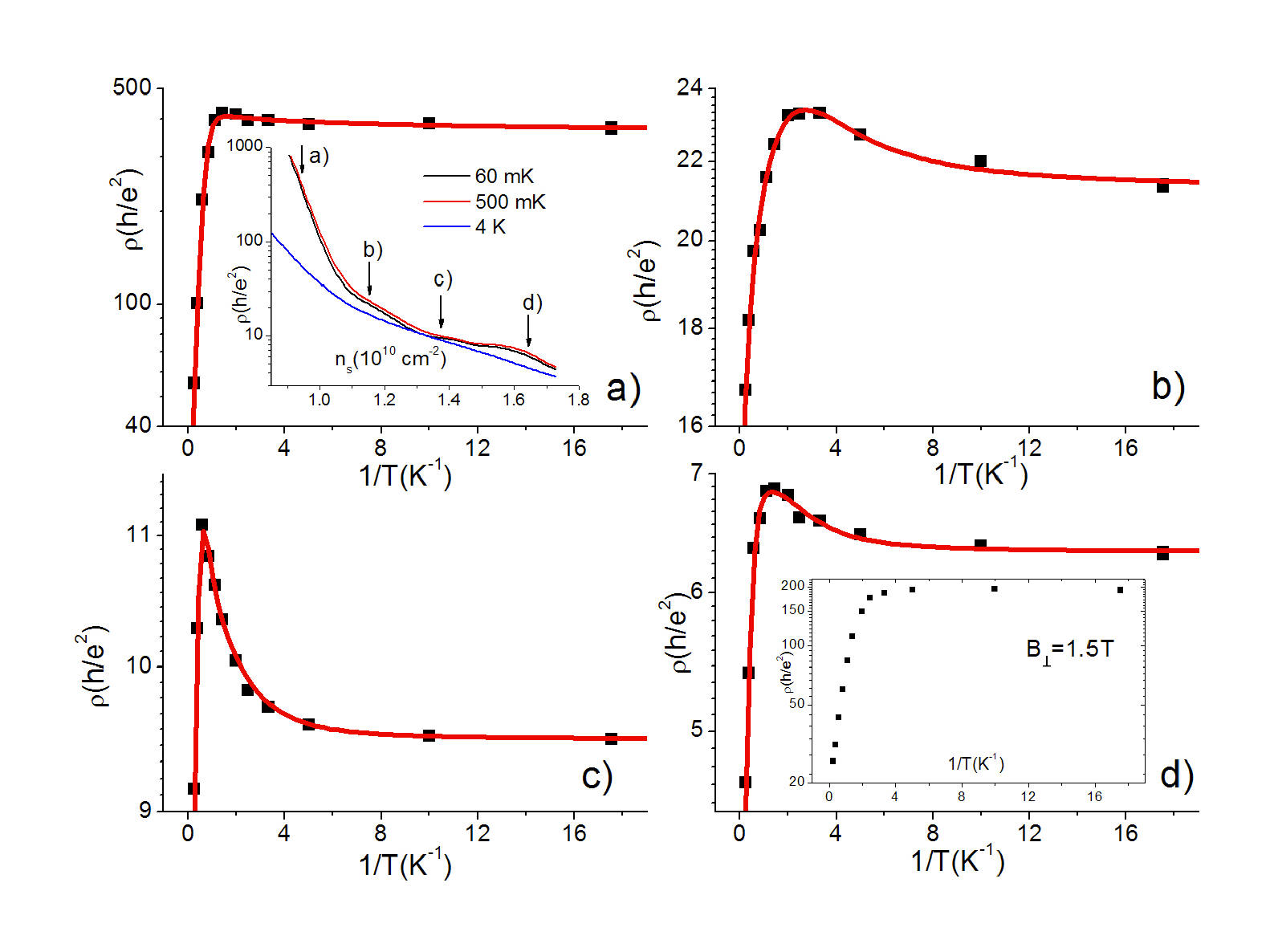}
\caption{(Colour online) Resistivity as a function of inverse temperature $1/T$ at $B=0$~T (symbols). At all densities, the strongly insulating $T$-dependence at higher temperatures is followed by a decrease in resistance at low $T$. Device dimensions are $W\times L=8~\mu\rm{m}\times0.5~\mu$m, spacer $\delta=40$~nm. Electron densities are indicated by arrows in the inset to a). Solid lines represent a fit of Eq.~\ref{eq1} to the data. Inset to a): Resistivity as a function of electron density at $T$=60~mK, 500~mK, 4~K. Inset to d): $\rho$ as function of $1/T$ at the same density as d) but at $B_{\perp}$=1.5T.}
\end{figure*}

\begin{figure*}[h]
\centering
\includegraphics[width=0.6\textwidth]{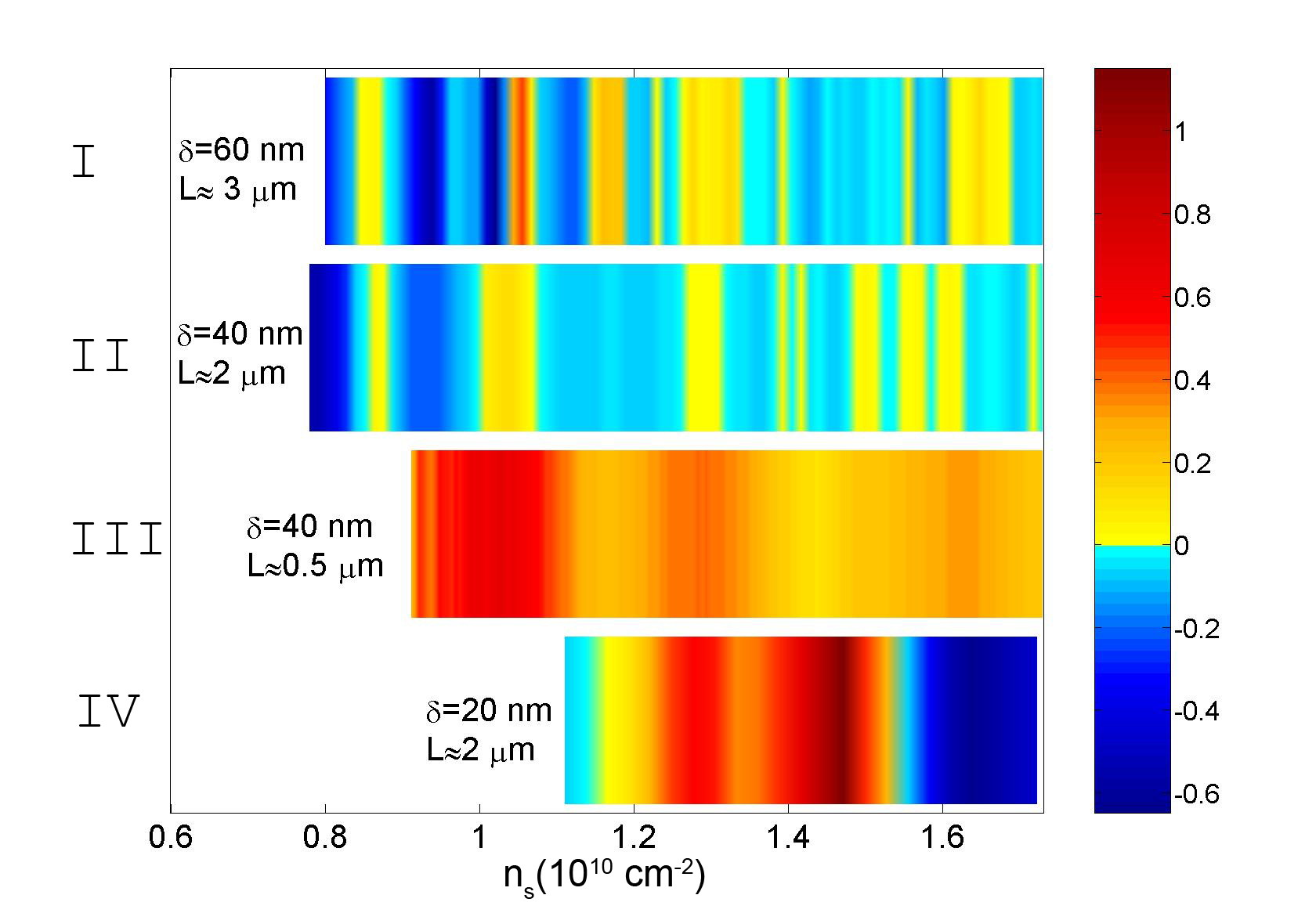}
\caption{(Colour online) Comparison of ``metallicity'' (defined as $\frac{d\rho(T)/dT}{\rho(T)}$ at $T=300$~mK) for four devices from three different wafers with spacer $\delta=20-60$ nm. A metallic state shows positive, and a saturated behaviour negative values. The mesa width is $W=8~\mu$m for all devices, while the gate lengths $L$ are indicated in the graph. Device III is shown in Fig.~2 and discussed extensively in the text.}
\end{figure*}

\begin{figure*}[h]
\centering
\includegraphics[width=0.6\textwidth]{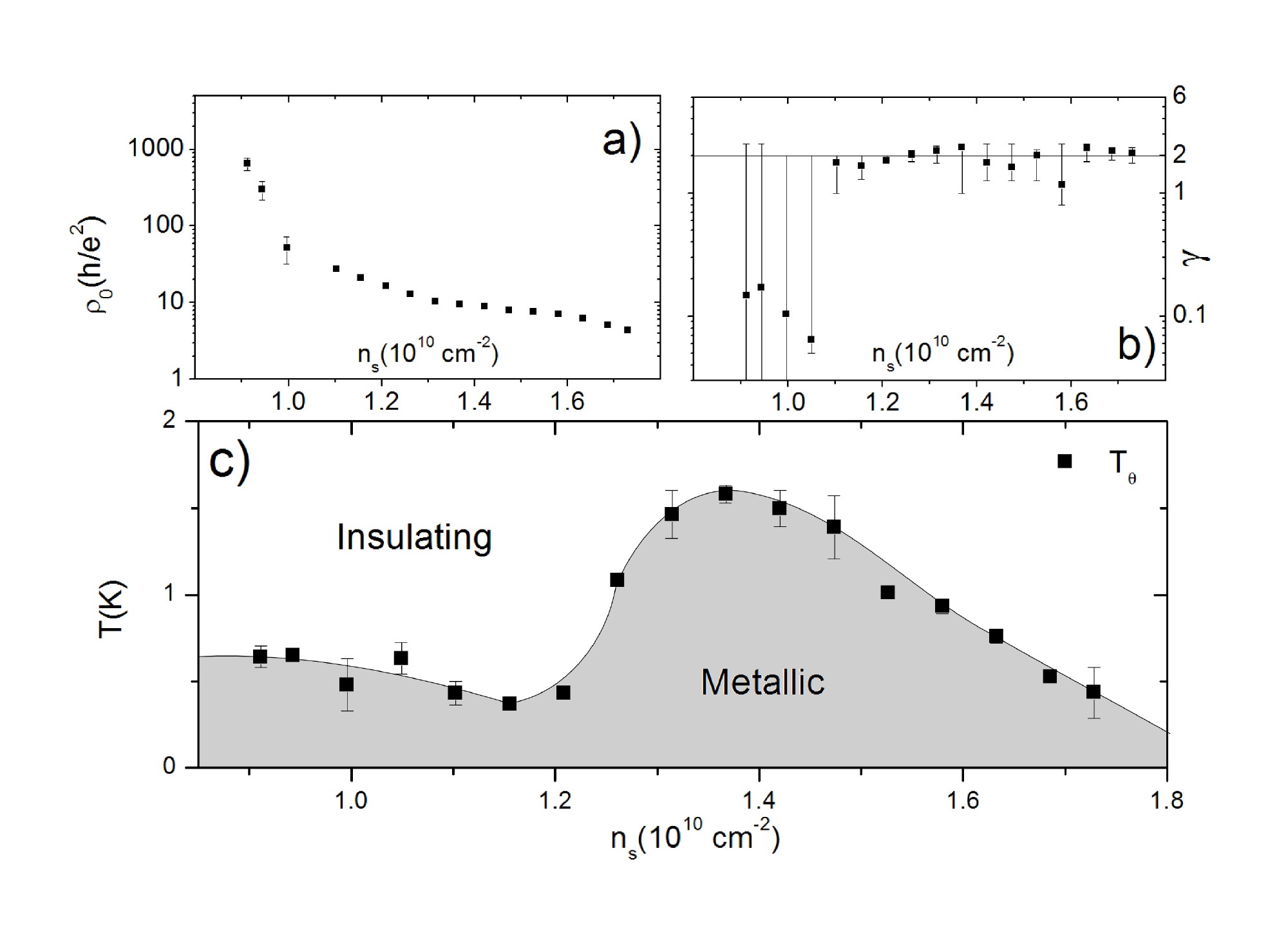}
\caption{Physical quantities extracted from data fits to Eq.~\ref{eq1}: a) $\rho_0$ is the resistivity extrapolated to zero $T$. Note that $\rho_0\gg h/e^2$ but finite for all $n_{\rm s}$. b) The temperature exponent is $\gamma\approx 2$ over a wide density range.
c) Temperature $T_{\theta}$ at which the resistivity is maximal. It shows a non-monotonic density-dependence with a maximum value $T_{\theta}\approx 1.6$~K. $T_{\theta}$ marks the transition between the insulating and the metallic phase, and, hence, defines a phase diagram in $T-n_{\rm s}$ space.}
\end{figure*}


\begin{thebibliography}{26}
\expandafter\ifx\csname natexlab\endcsname\relax\def\natexlab#1{#1}\fi
\expandafter\ifx\csname bibnamefont\endcsname\relax
  \def\bibnamefont#1{#1}\fi
\expandafter\ifx\csname bibfnamefont\endcsname\relax
  \def\bibfnamefont#1{#1}\fi
\expandafter\ifx\csname citenamefont\endcsname\relax
  \def\citenamefont#1{#1}\fi
\expandafter\ifx\csname url\endcsname\relax
  \def\url#1{\texttt{#1}}\fi
\expandafter\ifx\csname urlprefix\endcsname\relax\def\urlprefix{URL }\fi
\providecommand{\bibinfo}[2]{#2}
\providecommand{\eprint}[2][]{\url{#2}}

\bibitem[{\citenamefont{Mott et~al.}(1975)\citenamefont{Mott, Pepper, Pollitt, Wallis, and Adkins}}]{Mott1975}
\bibinfo{author}{\bibfnamefont{N.} \bibnamefont{Mott}},
\bibinfo{author}{\bibfnamefont{M.} \bibnamefont{Pepper}},
\bibinfo{author}{\bibfnamefont{S.} \bibnamefont{Pollitt}},
\bibinfo{author}{\bibfnamefont{R.~H.} \bibnamefont{Wallis}},
\bibinfo{author}{\bibfnamefont{C.~J.} \bibnamefont{Adkins}},
  \bibinfo{journal}{Proc. R. Soc. Lond. A.} \textbf{\bibinfo{volume}{345}},
  \bibinfo{pages}{169}  (\bibinfo{year}{1976}).

\bibitem[{\citenamefont{Abrahams et~al.}(1979)\citenamefont{Abrahams, Anderson,
  Licciardello, and Ramakrishnan}}]{Abrahams1979}
\bibinfo{author}{\bibfnamefont{E.}~\bibnamefont{Abrahams}},
  \bibinfo{author}{\bibfnamefont{P.~W.} \bibnamefont{Anderson}},
  \bibinfo{author}{\bibfnamefont{D.~C.} \bibnamefont{Licciardello}},
  \bibnamefont{and} \bibinfo{author}{\bibfnamefont{T.~V.}
  \bibnamefont{Ramakrishnan}}, \bibinfo{journal}{Phys. Rev. Lett.}
  \textbf{\bibinfo{volume}{42}}, \bibinfo{pages}{673} (\bibinfo{year}{1979}).

\bibitem[{\citenamefont{Abrahams et~al.}(2001)\citenamefont{Abrahams,
  Kravchenko, and Sarachik}}]{Abrahams2001}
\bibinfo{author}{\bibfnamefont{E.}~\bibnamefont{Abrahams}},
  \bibinfo{author}{\bibfnamefont{S.~V.} \bibnamefont{Kravchenko}},
  \bibnamefont{and} \bibinfo{author}{\bibfnamefont{M.~P.}
  \bibnamefont{Sarachik}}, \bibinfo{journal}{Rev. Mod. Phys.}
  \textbf{\bibinfo{volume}{73}}, \bibinfo{pages}{251} (\bibinfo{year}{2001}).
  
\bibitem[{\citenamefont{Fleishman et~al.}()\citenamefont{Fleishman, and Anderson}}]{Fleishman1980}
  \bibinfo{author}{\bibfnamefont{L.} \bibnamefont{Fleishman}}, \bibnamefont{and}
  \bibinfo{author}{\bibfnamefont{P.~W.} \bibnamefont{Anderson}},
  \bibinfo{journal}{Phys. Rev. B} \textbf{\bibinfo{volume}{21}},
  \bibinfo{pages}{2366} (\bibinfo{year}{1980}).  

\bibitem[{\citenamefont{Shepelyansky}(1994)}]{Shepelyansky1994}
\bibinfo{author}{\bibfnamefont{D.~L.} \bibnamefont{Shepelyansky}},
  \bibinfo{journal}{Phys. Rev. Lett.} \textbf{\bibinfo{volume}{73}},
  \bibinfo{pages}{2607} (\bibinfo{year}{1994}).

\bibitem[{\citenamefont{Basko et~al.}(2006)\citenamefont{Basko, Aleiner, and
  Altshuler}}]{Basko2006}
\bibinfo{author}{\bibfnamefont{D.~M.} \bibnamefont{Basko}},
  \bibinfo{author}{\bibfnamefont{I.~L.} \bibnamefont{Aleiner}},
  \bibnamefont{and} \bibinfo{author}{\bibfnamefont{B.~L.}
  \bibnamefont{Altshuler}}, \bibinfo{journal}{Cond-mat}
  \textbf{\bibinfo{volume}{0602510}}
  (\bibinfo{year}{2006}).

\bibitem[{\citenamefont{Benenti et~al.}(1999)\citenamefont{Benenti, Waintal,
  and Pichard}}]{Benenti1999}
\bibinfo{author}{\bibfnamefont{G.}~\bibnamefont{Benenti}},
  \bibinfo{author}{\bibfnamefont{X.}~\bibnamefont{Waintal}}, \bibnamefont{and}
  \bibinfo{author}{\bibfnamefont{J.-L.} \bibnamefont{Pichard}},
  \bibinfo{journal}{Phys. Rev. Lett.} \textbf{\bibinfo{volume}{83}},
  \bibinfo{pages}{1826} (\bibinfo{year}{1999}).

\bibitem[{\citenamefont{Katomeris et~al.}(2003)\citenamefont{Katomeris, Selva,
  and Pichard}}]{Katomeris2003}
\bibinfo{author}{\bibfnamefont{G.}~\bibnamefont{Katomeris}},
  \bibinfo{author}{\bibfnamefont{F.}~\bibnamefont{Selva}}, \bibnamefont{and}
  \bibinfo{author}{\bibfnamefont{J.-L.} \bibnamefont{Pichard}},
  \bibinfo{journal}{Eur. Phys. J. B} \textbf{\bibinfo{volume}{31}},
  \bibinfo{pages}{401} (\bibinfo{year}{2003}).
  
\bibitem[{\citenamefont{Henseler et~al.}(2007)\citenamefont{Henseler, Kroha,
  and Shapiro}}]{Henseler2007}
\bibinfo{author}{\bibfnamefont{P.} \bibnamefont{Henseler}},
\bibinfo{author}{\bibfnamefont{J.} \bibnamefont{Kroha}},
\bibinfo{author}{\bibfnamefont{B.} \bibnamefont{Shapiro}},
  \bibinfo{journal}{Cond-mat} \textbf{\bibinfo{volume}{0707.1301v1}}
  (\bibinfo{year}{2007}).
  
\bibitem[{\citenamefont{Efros and Shklovskii}(1975)}]{Efros1975}
\bibinfo{author}{\bibfnamefont{A.~L.} \bibnamefont{Efros}} \bibnamefont{and}
  \bibinfo{author}{\bibfnamefont{B.~I.} \bibnamefont{Shklovskii}},
  \bibinfo{journal}{J. Phys. C} \textbf{\bibinfo{volume}{8}},
  \bibinfo{pages}{L49} (\bibinfo{year}{1975}).

\bibitem[{\citenamefont{Tanatar and Ceperley}(1989)}]{Tanatar1989}
\bibinfo{author}{\bibfnamefont{B.}~\bibnamefont{Tanatar}} \bibnamefont{and}
  \bibinfo{author}{\bibfnamefont{D.~M.} \bibnamefont{Ceperley}},
  \bibinfo{journal}{Phys. Rev. B} \textbf{\bibinfo{volume}{39}},
  \bibinfo{pages}{5005} (\bibinfo{year}{1989}).

\bibitem[{\citenamefont{Normand et~al.}(1992)\citenamefont{Normand, Littlewood,
  and Millis}}]{Normand1992}
\bibinfo{author}{\bibfnamefont{B.~G.~A.} \bibnamefont{Normand}},
  \bibinfo{author}{\bibfnamefont{P.~B.} \bibnamefont{Littlewood}},
  \bibnamefont{and} \bibinfo{author}{\bibfnamefont{A.~J.}
  \bibnamefont{Millis}}, \bibinfo{journal}{Phys. Rev. B}
  \textbf{\bibinfo{volume}{46}}, \bibinfo{pages}{3920} (\bibinfo{year}{1992}).

\bibitem[{\citenamefont{Sarma et~al.}(2005)\citenamefont{Sarma, Lilly, Hwang,
  Pfeiffer, West, and Reno}}]{Sarma2005}
\bibinfo{author}{\bibfnamefont{S.} \bibnamefont{Das Sarma}},
  \bibinfo{author}{\bibfnamefont{M.~P.} \bibnamefont{Lilly}},
  \bibinfo{author}{\bibfnamefont{E.~H.} \bibnamefont{Hwang}},
  \bibinfo{author}{\bibfnamefont{L.~N.} \bibnamefont{Pfeiffer}},
  \bibinfo{author}{\bibfnamefont{K.~W.} \bibnamefont{West}}, \bibnamefont{and}
  \bibinfo{author}{\bibfnamefont{J.~L.} \bibnamefont{Reno}},
  \bibinfo{journal}{Phys. Rev. Lett.} \textbf{\bibinfo{volume}{94}},
  \bibinfo{pages}{136401} (\bibinfo{year}{2005}).

\bibitem[{\citenamefont{Efros}(1988)}]{Efros1988}
\bibinfo{author}{\bibfnamefont{A.~L.} \bibnamefont{Efros}},
  \bibinfo{journal}{Solid State Comm.} \textbf{\bibinfo{volume}{65}},
  \bibinfo{pages}{1281} (\bibinfo{year}{1988}).
  

\bibitem[{\citenamefont{Finkelstein et~al.}(2000)\citenamefont{Finkelstein,
  Glicofridis, Ashoori, and Shayegan}}]{Finkelstein2000}
\bibinfo{author}{\bibfnamefont{G.}~\bibnamefont{Finkelstein}},
  \bibinfo{author}{\bibfnamefont{P.~I.} \bibnamefont{Glicofridis}},
  \bibinfo{author}{\bibfnamefont{R.~C.} \bibnamefont{Ashoori}},
  \bibnamefont{and} \bibinfo{author}{\bibfnamefont{M.}~\bibnamefont{Shayegan}},
  \bibinfo{journal}{Science} \textbf{\bibinfo{volume}{289}},
  \bibinfo{pages}{90} (\bibinfo{year}{2000}).

\bibitem[{\citenamefont{Chakraborty et~al.}(2004)\citenamefont{Chakraborty,
  Maasilta, Tessmer, and Melloch}}]{Chakraborty2004}
\bibinfo{author}{\bibfnamefont{S.}~\bibnamefont{Chakraborty}},
  \bibinfo{author}{\bibfnamefont{I.~J.} \bibnamefont{Maasilta}},
  \bibinfo{author}{\bibfnamefont{S.~H.} \bibnamefont{Tessmer}},
  \bibnamefont{and} \bibinfo{author}{\bibfnamefont{M.~R.}
  \bibnamefont{Melloch}}, \bibinfo{journal}{Phys. Rev. B}
  \textbf{\bibinfo{volume}{69}}, \bibinfo{pages}{073308}
  (\bibinfo{year}{2004}).

\bibitem[{\citenamefont{Baenninger et~al.}(2005)\citenamefont{Baenninger,
  Ghosh, Pepper, Beere, Farrer, Atkinson, and Ritchie}}]{Baenninger2005}
\bibinfo{author}{\bibfnamefont{M.}~\bibnamefont{Baenninger}},
  \bibinfo{author}{\bibfnamefont{A.}~\bibnamefont{Ghosh}},
  \bibinfo{author}{\bibfnamefont{M.}~\bibnamefont{Pepper}},
  \bibinfo{author}{\bibfnamefont{H.~E.} \bibnamefont{Beere}},
  \bibinfo{author}{\bibfnamefont{I.}~\bibnamefont{Farrer}},
  \bibinfo{author}{\bibfnamefont{P.}~\bibnamefont{Atkinson}}, \bibnamefont{and}
  \bibinfo{author}{\bibfnamefont{D.~A.} \bibnamefont{Ritchie}},
  \bibinfo{journal}{Phys. Rev. B} \textbf{\bibinfo{volume}{72}},
  \bibinfo{pages}{241311(R)} (\bibinfo{year}{2005}).

\bibitem[{\citenamefont{Haviland et~al.}(1989)\citenamefont{Haviland, Liu, and
  Goldman}}]{Haviland1989}
\bibinfo{author}{\bibfnamefont{D.~B.} \bibnamefont{Haviland}},
  \bibinfo{author}{\bibfnamefont{Y.}~\bibnamefont{Liu}}, \bibnamefont{and}
  \bibinfo{author}{\bibfnamefont{A.~M.} \bibnamefont{Goldman}},
  \bibinfo{journal}{Phys. Rev. Lett.} \textbf{\bibinfo{volume}{62}},
  \bibinfo{pages}{2180} (\bibinfo{year}{1989}).

\bibitem[{\citenamefont{Shahar et~al.}(1995)\citenamefont{Shahar, Tsui,
  Shayegan, Bhatt, and Cunningham}}]{Shahar1995}
\bibinfo{author}{\bibfnamefont{D.}~\bibnamefont{Shahar}},
  \bibinfo{author}{\bibfnamefont{D.~C.} \bibnamefont{Tsui}},
  \bibinfo{author}{\bibfnamefont{M.}~\bibnamefont{Shayegan}},
  \bibinfo{author}{\bibfnamefont{R.~N.} \bibnamefont{Bhatt}}, \bibnamefont{and}
  \bibinfo{author}{\bibfnamefont{J.~E.} \bibnamefont{Cunningham}},
  \bibinfo{journal}{Phys. Rev. Lett.} \textbf{\bibinfo{volume}{74}},
  \bibinfo{pages}{4511} (\bibinfo{year}{1995}).

\bibitem[{\citenamefont{Simmons et~al.}(1998)\citenamefont{Simmons, Hamilton,
  Pepper, Linfield, Rose, Ritchie, Savchenko, and Griffiths}}]{Simmons1998}
\bibinfo{author}{\bibfnamefont{M.~Y.} \bibnamefont{Simmons}},
  \bibinfo{author}{\bibfnamefont{A.~R.} \bibnamefont{Hamilton}},
  \bibinfo{author}{\bibfnamefont{M.}~\bibnamefont{Pepper}},
  \bibinfo{author}{\bibfnamefont{E.~H.} \bibnamefont{Linfield}},
  \bibinfo{author}{\bibfnamefont{P.~D.} \bibnamefont{Rose}},
  \bibinfo{author}{\bibfnamefont{D.~A.} \bibnamefont{Ritchie}},
  \bibinfo{author}{\bibfnamefont{A.~K.} \bibnamefont{Savchenko}},
  \bibnamefont{and} \bibinfo{author}{\bibfnamefont{T.~G.}
  \bibnamefont{Griffiths}}, \bibinfo{journal}{Phys. Rev. Lett.}
  \textbf{\bibinfo{volume}{80}}, \bibinfo{pages}{1292} (\bibinfo{year}{1998}).

\bibitem[{\citenamefont{Ghosh et~al.}(2004{\natexlab{a}})\citenamefont{Ghosh,
  Pepper, Beere, and Ritchie}}]{Ghosh2004a}
\bibinfo{author}{\bibfnamefont{A.}~\bibnamefont{Ghosh}},
  \bibinfo{author}{\bibfnamefont{M.}~\bibnamefont{Pepper}},
  \bibinfo{author}{\bibfnamefont{H.~E.} \bibnamefont{Beere}}, \bibnamefont{and}
  \bibinfo{author}{\bibfnamefont{D.~A.} \bibnamefont{Ritchie}},
  \bibinfo{journal}{J. Phys. C} \textbf{\bibinfo{volume}{16}},
  \bibinfo{pages}{3623} (\bibinfo{year}{2004}{\natexlab{a}}).

\bibitem[{\citenamefont{Tripathi and Kennett}(2006)}]{Tripathi2006}
\bibinfo{author}{\bibfnamefont{V.}~\bibnamefont{Tripathi}} \bibnamefont{and}
  \bibinfo{author}{\bibfnamefont{M.~P.} \bibnamefont{Kennett}},
  \bibinfo{journal}{Phys. Rev. B} \textbf{\bibinfo{volume}{74}},
  \bibinfo{pages}{195334} (\bibinfo{year}{2006}).

\bibitem[{\citenamefont{Slutskin et~al.}(2000)\citenamefont{Slutskin, Slavin,
  and Kovtun}}]{Slutskin2000}
\bibinfo{author}{\bibfnamefont{A.~A.} \bibnamefont{Slutskin}},
  \bibinfo{author}{\bibfnamefont{V.~V.} \bibnamefont{Slavin}},
  \bibnamefont{and} \bibinfo{author}{\bibfnamefont{H.~A.}
  \bibnamefont{Kovtun}}, \bibinfo{journal}{Phys. Rev. B}
  \textbf{\bibinfo{volume}{61}}, \bibinfo{pages}{14184} (\bibinfo{year}{2000}).

\bibitem[{\citenamefont{Andreev and Lifshitz}(1969)}]{Andreev1969}
\bibinfo{author}{\bibfnamefont{A.~F.} \bibnamefont{Andreev}} \bibnamefont{and}
  \bibinfo{author}{\bibfnamefont{I.~M.} \bibnamefont{Lifshitz}},
  \bibinfo{journal}{Sov. Phys. JETP} \textbf{\bibinfo{volume}{48}},
  \bibinfo{pages}{1107} (\bibinfo{year}{1969}).

\bibitem[{\citenamefont{Shapiro}(2001)}]{Shapiro2001}
\bibinfo{author}{\bibfnamefont{B.}~\bibnamefont{Shapiro}},
  \bibinfo{journal}{Phil. Mag. B} \textbf{\bibinfo{volume}{81}},
  \bibinfo{pages}{1301} (\bibinfo{year}{2001}).

\bibitem[{\citenamefont{Spivak}(2003)}]{Spivak2003}
\bibinfo{author}{\bibfnamefont{B.}~\bibnamefont{Spivak}},
  \bibinfo{journal}{Phys. Rev. B} \textbf{\bibinfo{volume}{67}},
  \bibinfo{pages}{125205} (\bibinfo{year}{2003}).
  
\bibitem[{\citenamefont{Chui and Tanatar}(1995)}]{Chui1995}
\bibinfo{author}{\bibfnamefont{S.~T.} \bibnamefont{Chui}} \bibnamefont{and}
  \bibinfo{author}{\bibfnamefont{B.}~\bibnamefont{Tanatar}},
  \bibinfo{journal}{Phys. Rev. Lett.} \textbf{\bibinfo{volume}{74}},
  \bibinfo{pages}{458} (\bibinfo{year}{1995}).

\bibitem[{\citenamefont{Ghosh et~al.}(2004{\natexlab{b}})\citenamefont{Ghosh,
  Pepper, Beere, and Ritchie}}]{Ghosh2004}
\bibinfo{author}{\bibfnamefont{A.}~\bibnamefont{Ghosh}},
  \bibinfo{author}{\bibfnamefont{M.}~\bibnamefont{Pepper}},
  \bibinfo{author}{\bibfnamefont{H.~E.} \bibnamefont{Beere}}, \bibnamefont{and}
  \bibinfo{author}{\bibfnamefont{D.~A.} \bibnamefont{Ritchie}},
  \bibinfo{journal}{Phys. Rev. B} \textbf{\bibinfo{volume}{70}},
  \bibinfo{pages}{233309} (\bibinfo{year}{2004}{\natexlab{b}}).

\bibitem[{\citenamefont{Pushkarov}(2003)}]{Pushkarov2003}
\bibinfo{author}{\bibfnamefont{D.~I.} \bibnamefont{Pushkarov}},
  \bibinfo{journal}{Cond-mat} \textbf{\bibinfo{volume}{0310283}}
  (\bibinfo{year}{2003}).

\bibitem[{\citenamefont{Simmons et~al.}(2000)\citenamefont{Simmons, Hamilton,
  Pepper, Linfield, Rose, and Ritchie}}]{Simmons2000}
\bibinfo{author}{\bibfnamefont{M.~Y.} \bibnamefont{Simmons}},
  \bibinfo{author}{\bibfnamefont{A.~R.} \bibnamefont{Hamilton}},
  \bibinfo{author}{\bibfnamefont{M.}~\bibnamefont{Pepper}},
  \bibinfo{author}{\bibfnamefont{E.~H.} \bibnamefont{Linfield}},
  \bibinfo{author}{\bibfnamefont{P.~D.} \bibnamefont{Rose}}, \bibnamefont{and}
  \bibinfo{author}{\bibfnamefont{D.~A.} \bibnamefont{Ritchie}},
  \bibinfo{journal}{Phys. Rev. Lett.} \textbf{\bibinfo{volume}{84}},
  \bibinfo{pages}{2489} (\bibinfo{year}{2000}).

\end{thebibliography}
\end{document}